# COSMIC MICROWAVE BACKGROUND ANISOTROPIES AND THE GEOMETRY OF THE UNIVERSE

MARC KAMIONKOWSKI *

*School of Natural Sciences*
*Institute for Advanced Study*
*Princeton, NJ 08540, USA*

ABSTRACT

In this talk, I review some recent work on cosmic microwave background (CMB) anisotropies in an open universe. I emphasize that the observed CMB anisotropies are still consistent with a low value of $\Omega$, and I address the question of whether future CMB measurements will be able to provide information on the geometry of the Universe.

## 1. Introduction

Determination of the geometry of the Universe remains perhaps the most important goal of cosmology. Even the man on the street wants to know if the Universe will re-collapse or expand forever. There is a tremendous theoretical prejudice for a flat Universe, but the observational situation remains unclear. Some measurements are consistent with a flat Universe, yet numerous others suggest the Universe is open. Even so, the vast majority of theoretical work on cosmic microwave background (CMB) anisotropies and the evolution of structure in the Universe has been performed assuming the Universe is flat.

In this talk, I review recent work done in collaboration with David Spergel, Naoshi Sugiyama, and Bharat Ratra on cosmic microwave background anisotropies in an open Universe[1,2,3], and address the question of whether CMB anisotropies can be used to determine the geometry of the Universe. To clarify, I will take $\Omega$ to be the *total* (i.e., matter, radiation, and vacuum energy or cosmological constant) mass density of the Universe in units of critical density, so low $\Omega$ implies an open Universe. First I will argue that it is possible for COBE-normalized low-$\Omega$ models to be consistent with observed large-scale structure[1,4]. I then consider whether information on the value of $\Omega$ is encoded in large-angle and/or small-angle anisotropies[1,2].

In the next Section, I will discuss open-Universe models with scale-invariant primordial adiabatic density density perturbations and cold dark matter and ask whether large-angle anisotropies probe the geometry of the Universe[1]. In Section 3, I argue that small-scale anisotropies can potentially be used to determine the geometry of the Universe[2], and I conclude in the final Section.

## 2. Standard CDM in an Open Universe

The most studied and perhaps best-motivated model for the origin of structure in a *flat* Universe is the standard CDM model. In standard CDM, there is a flat, scale-invariant spectrum of primordial density perturbations, produced by inflation, and the dark matter is cold. To a large extent, the fit of this simple model to the data is quite impressive; however, there are some flaws. One of these problems, for example, comes from the power spectrum on large scales.

In standard CDM, the primordial power spectrum is flat, $P(k) \propto k^n$ with $n = 1$. Sub-horizon sized perturbations grow when the Universe is matter dominated, but not when it is radiation dominated. Therefore, the growth of density perturbations on small distance scales which come into the horizon before radiation domination is stunted relative to those on the larger scales which come into the horizon after the Universe is matter dominated. Therefore, the observed, or processed, power spectrum is $P(k) \propto k$ at large scales and $P(k) \propto k^{-3}$ at small scales. The scale at which the power spectrum turns over measures the size of the horizon at matter-radiation equality, and this depends on the combination $\Omega h$, where $h$ is the Hubble parameter in units of 100 km sec$^{-1}$ Mpc$^{-1}$. Although the shape of the power spectrum measured by the APM, IRAS, and CfA surveys is described well by the standard-CDM transfer function, it seems that the break in the power spectrum is best fit by $\Omega h \simeq 0.25$ (Ref. 5). This cannot be reconciled with a flat Universe, even if $h$ is as low as 0.4. The data are shown in Fig. 1.

However, the standard-CDM transfer function can fit the data quite well if $\Omega$ is low. In recent work, we considered whether the amplitude of the power spectrum in such a low-$\Omega$ model is consistent when normalized to COBE[1]. The main problem is to come up with an open-universe analog of a scale-invariant spectrum in the low-$\Omega$ variant of the standard-CDM model. Unlike a flat universe, which is truly scale free, an open universe has a scale in it: the curvature scale. On scales much smaller than the curvature scale, the spectrum should be flat, $P(k) \propto k$, but the extrapolation to scales larger than the curvature scale is ambiguous. To address this problem, we made several *ansatzes* for the power spectrum on super-curvature sized scales, and then checked the dependence of the numerical results for COBE normalization of the spectrum on the various *ansatzes*.

In a flat universe, a power law in distance is also a power law in volume as well as a power law in eigenvalue of the Laplace operator. In an open universe, power laws in distance, volume, and eigenvalue of the Laplace operator are distinct. Therefore, we took power laws in these three quantities as our three *ansatzes* for the power spectra. Such power spectra could conceivably have arisen if the Universe underwent a period of "frustrated" inflation. It turns out that at least in a specific realization of such an inflationary model, the power spectrum looks significantly different on super-curvature scales[6], so we include this as well. In Fig. 2, we show the processed power spectra for our three *ansatzes* and for the inflationary model as well. On scales smaller than the curvature scale, the power spectra are all similar, as they should be, but on scales larger than the curvature scale, they are dramatically different, especially the inflationary spectrum. The volume scaling provides the least power on large scales, and the inflationary model provides the most.

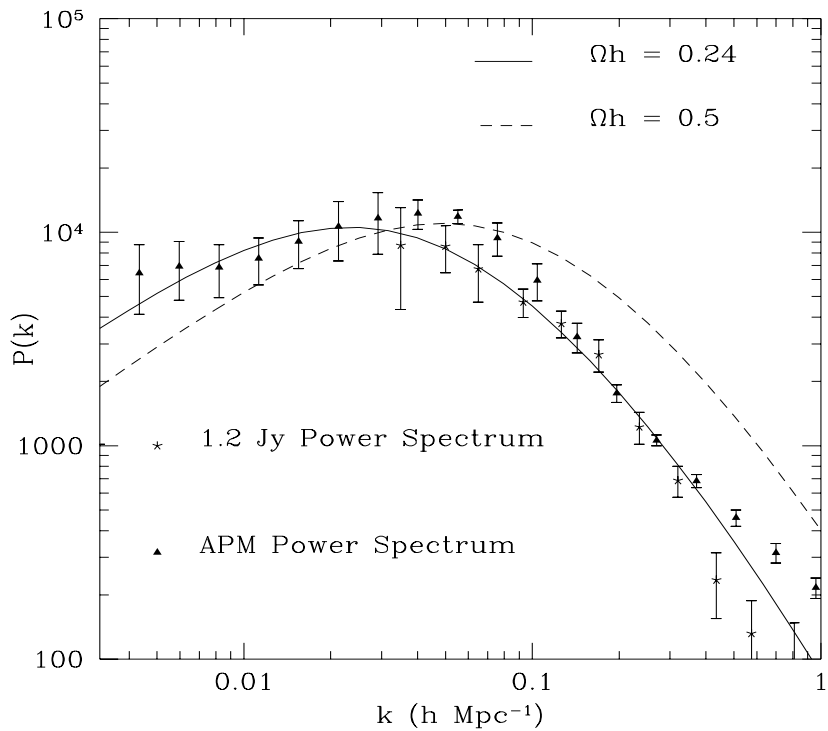

FIG. 1. Processed scale-invariant spectrum with $\Omega h = 0.24$ and $\Omega h = 0.5$. From Ref. 1.

Given a specific functional form for the spectrum of primordial density perturbations, it is straightforward to calculate the resulting microwave anisotropies[1]. One can use the COBE detection of CMB anisotropies to normalize the models and compare with the observed structure on small scales. Fig. 3 shows the results for the power spectrum that scales with volume. We plot contours of the age of the Universe, $t_U = 10$ and $13$ Gyrs. The shaded regions are those where $0.2 < \Omega h < 0.3$ as suggested by the APM, IRAS, and CfA surveys. Also plotted are contours of $\sigma_8^{\mathrm{mass}}$, the variance of the mass distribution on the $8h^{-1}$-Mpc scale. If optical galaxies trace the mass distribution, then $\sigma_8^{\mathrm{mass}} \simeq 1$. However, the distribution of infrared galaxies is (anti)biased relative to the optical galaxies[7]; if they trace the mass, then $\sigma_8^{\mathrm{mass}} \simeq 0.7$. If we are willing to allow a bias of a factor of two of optical galaxies relative to the mass, then $\sigma_8^{\mathrm{mass}}$ could be as low as $0.5$. If so, then Fig. 3 shows that low-$\Omega$ standard-CDM models with $0.4 \lesssim \Omega \lesssim 0.8$ are consistent with the power spectrum measured by the APM, IRAS, and CfA surveys and with the age of the Universe. If we demand consistency with large-scale peculiar velocity flows, which suggest a large value of $\Omega$ (Ref. 8), some of the available parameter space will be eliminated, although it should be noted that a different analysis with different data may allow for values of $\Omega$ smaller than previously believed[9].

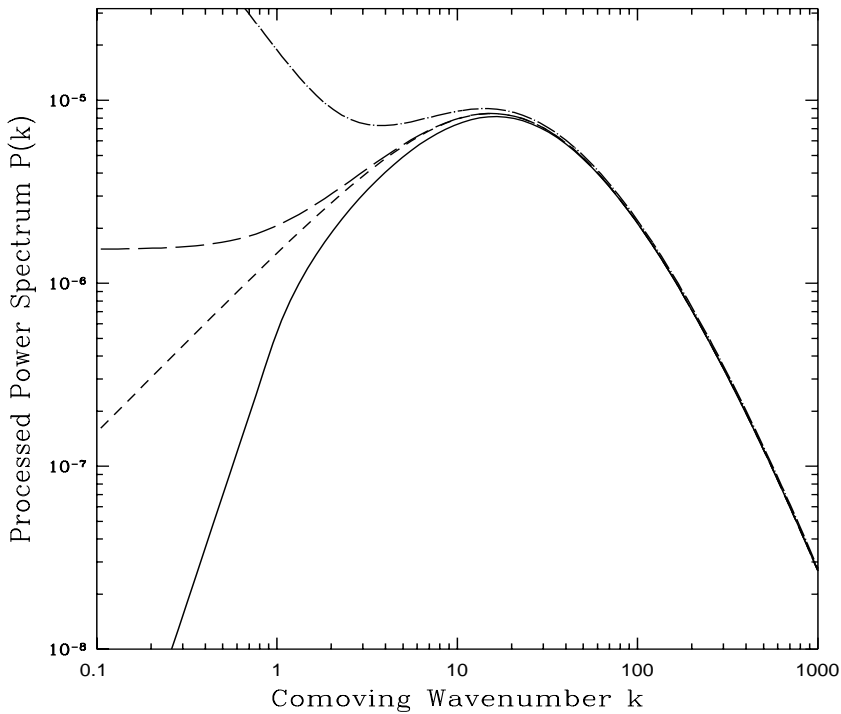

FIG. 2. Processed power spectra for primordial spectra that are power laws in volume (solid curve), wavelength (short-dash curve), and eigenvalue of the Laplace operator (long-dash curve). Also shown is the inflationary spectrum (dot-dash curve). From Ref. 1.

In Fig. 3, we showed the results for the volume scaling shown in Fig. 2. However, from the looks of Fig. 2, it seems that we would have obtained entirely different results if we had used one of the other spectra, such as the inflationary spectrum. It turns out that the basic conclusions illustrated in Fig. 3 are the same even if we use the inflationary spectrum. This is easily understood. If the Universe is open, then the angular scale subtended by the curvature scale at the redshift $z \simeq 1100$ at which CMB photons last scattered is roughly $[\Omega/(1-\Omega)]^{1/2}$, which is about $20°$ if $\Omega \simeq 0.1$ and $40°$ if $\Omega \simeq 0.3$. So, if $\Omega \simeq 0.1$, only multipole moments $l \lesssim 9$ probe scales larger than the curvature scale, and if $\Omega \simeq 0.3$, only multipole moments $l \lesssim 4$ probe scales larger than the curvature scale. Therefore, only the lowest CMB moments will differ appreciably if only the super-curvature scale perturbations are different. Furthermore, the lowest multipole moments are subject to the largest theoretical uncertainty due to cosmic variance, so they contribute little to the total COBE normalization. So the discrepancy in the COBE normalization between the various *ansatzes* is relatively small, and it becomes smaller for larger $\Omega$.

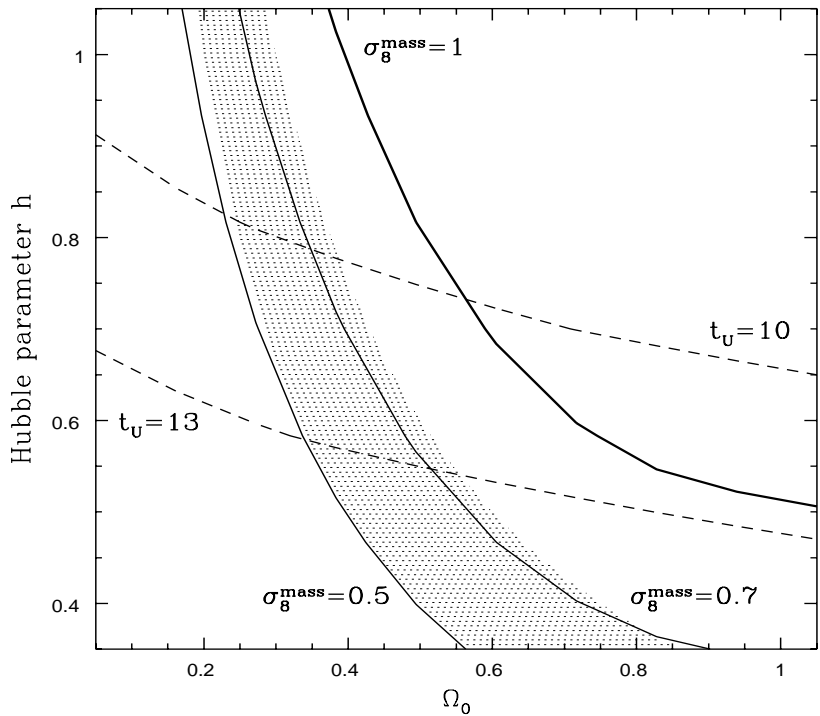

FIG. 3. Contour plot of $\sigma_8^{\mathrm{mass}}$ obtained from COBE normalization in the $\Omega$-$h$ plane. The heavy solid curve is the contour of $\sigma_8^{\mathrm{mass}} = 1$; the upper and lower lighter solid curves are contours of $\sigma_8^{\mathrm{mass}} = 0.7$ and 0.5, respectively. The upper and lower broken curves are contours of age of the universe of 10 and 13 Gyrs respectively. The shaded region is that where $0.2 \leq \Omega h \leq 0.3$, as suggested by the observed power on large scales. From Ref. 1.

Qualitatively, if we COBE normalize the inflationary power spectrum, rather than the volume scaling, then the biasing required to fit the observed power spectrum on small scales is larger; i.e., we would predict a smaller $\sigma_8^{\mathrm{mass}}$. Numerically, the difference in the normalization is quite sizable for very low $\Omega$; for $\Omega = 0.1$, the bias required when normalizing the inflationary spectrum is about a factor of two larger than that required when normalizing the volume scaling. Although this is quite a substantial difference, it is irrelevant to the final conclusions of Ref. 1, since such a low-density model requires too much biasing to be acceptable. For $\Omega = 0.3$, the difference in normalization between the two types of spectra is about 30%, and the difference is even smaller for larger $\Omega$. Therefore, the main result shown in Fig. 3—that a COBE-normalized standard-CDM type model with $0.4 \lesssim \Omega \lesssim 0.8$ is consistent with the observed power spectrum—will be valid for the inflationary power spectrum as well[3].

Now consider the question of whether large-angle anisotropies can be used to measure $\Omega$. The ambiguity in the super-curvature spectra shown in Fig. 2 suggests that the answer is "no." Although there may indeed be a feature in the CMB-anisotropy spectrum at the curvature scale, our uncertainty about the primordial spectrum does not allow us to make any specific prediction for the shape of the anisotropy spectrum for any given value of $\Omega$. Moreover, the biggest difference in the CMB spectrum in an open universe would be at the lowest multipole moments. Due to cosmic variance, these observed quantities taken alone cannot reliably discriminate between models. On the other hand, deviations from a standard scale-invariant spectrum on the largest angular scales may be at least suggestive of a low-density universe, or some other deviation from standard CDM, although such observations are by no means conclusive when taken alone.

## 3. Small-Angle Anisotropies as a Probe of $\Omega$

I now briefly review recent work where it was proposed that small-angle anisotropies could potentially be used to determine the geometry of the Universe[2]. The basic idea is simple: The horizon at the surface of last scatter subtends an angle of about $\Omega^{1/2}$ degrees on the sky. This is simply a geometric effect. There is more volume for a given distance at large redshift in an open universe than in a flat universe. In the CMB-anisotropy spectrum, the horizon at the surface of last scatter is marked by the angular location of the Doppler peak. In Ref. 2, we argued that this conclusion is quite insensitive to assumptions about the values of $h$, the baryon density, the cosmological constant, the spectral index, the contribution from gravitational waves, and the ionization history (as long as there was not enough reionization to fully erase the Doppler peak). These effects may raise or lower the Doppler peak, but to a great extent, the angle at which the first Doppler peak occurs depends only on $\Omega$. Therefore, the angular location of the Doppler peak provides a measure of the value of $\Omega$.

In Fig. 4, we show the results of numerical computations of the CMB spectrum as a function of multipole moment $l$ for flat scale-invariant spectra for several values of $\Omega$. We show results both for models with no reionization $\tau = 0$, and reionized models where the optical depth to the surface of last scatter is $\tau = 1$. Fig. 4 show that our simple heuristic arguments about the location of the Doppler peak are borne out by detailed numerical calculations. Additional numerical results[2,10] show that the location of the Doppler peak is indeed highly insensitive to the other undetermined cosmological parameters. It still remains to be seen, however, if the anisotropies can be mapped precisely enough to determine $\Omega$.

If, for example, reionization erases the Doppler peak, then this test will not work. However, it can be argued that in these models, the optical depth to the surface of last scatter should be $\tau \lesssim 1$. Therefore, the Doppler peak should still be distinguishable even if it is suppressed[2]. Similarly, if the primordial density perturbations are not adiabatic, then the degree-scale anisotropies could be markedly different. It could be argued that if the Universe is indeed open, then there is no well-motivated mechanism for producing a primordial standard-CDM type perturbation spectrum, in which case our technique could not be used to measure $\Omega$. However, the basic idea could still apply: Causal mechanisms at the surface of last scatter work only on scales smaller than the horizon, so any features in the CMB spectrum should occur on angular scales which subtend distance scales smaller than the horizon.

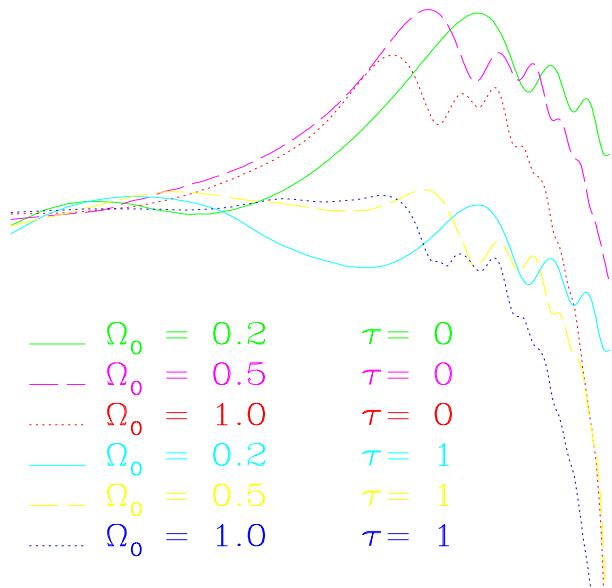

FIG. 4. The COBE-normalized CMB spectrum as a function of multipole moment $l$ for several values of $\Omega$ and for optical depths $\tau = 0$ (no reionization) and $\tau = 1$. Here we have taken $\Omega_b = 0.06$, $h = 0.5$, and $\Lambda = 0$. From Ref. 2.

In particular, if CMB experiments accurately measure the standard-CDM power spectrum and find a Doppler peak at a degree, and not at a smaller angle, it provides very strong evidence that the Universe is indeed flat (or at least not open). In an open universe, this angular scale is outside the horizon at decoupling, so there is no way the standard $\Omega = 1$ Doppler peak could be mimicked without invoking some contrived and pathological spectrum of primordial density perturbations.

## 4. Conclusions

Since the COBE discovery of anisotropies in the CMB, theorists have interpreted the observations primarily within the context of a flat universe. The standard lore claimed that the observed anisotropies were too small to account for the observed structure in an open universe, and that the absence of a dramatic feature in the spectrum at large angles was evidence for a flat universe. In this talk, I have argued that neither of these statements need be true. The current CMB measurements are consistent with an open Universe, and the value of $\Omega$ has yet to be determined.

Although current observational results do not tell us about the value of $\Omega$, there is a realistic possibility that forthcoming small-angle CMB measurements will be able to tell us if the Universe is flat. The angular location of the Doppler peak depends on the geometry of the Universe. In particular, it is almost inconceivable that the standard $\Omega = 1$ Doppler peak could arise at the proper angular scale in an open Universe. The challenge for the forthcoming CMB experiments will be to map the location of the Doppler peak with enough precision to make the measurement of $\Omega$ feasible.

The work reported here was done with Bharat Ratra, David Spergel, and Naoshi Sugiyama, and I thank them for very fruitful and enjoyable collaborations. This work was supported by the U.S. Department of Energy under contract DEFG02-90-ER 40542.